# Hybrid Group IV Nanophotonic Structures Incorporating Diamond Silicon-Vacancy Color Centers

*Jingyuan Linda Zhang[‡1], Hitoshi Ishiwata[‡2,3], Thomas M. Babinec[1], Marina Radulaski[1], Kai Müller[1], Konstantinos G. Lagoudakis[1], Constantin Dory[1], Jeremy Dahl[2], Robert Edgington[2], Veronique Soulière[4], Gabriel Ferro[4], Andrey A. Fokin[5], Peter R. Schreiner[5], Zhi-Xun Shen[2,3], Nicholas A. Melosh[2,3], Jelena Vučković[1]*

[1]E. L. Ginzton Laboratory, Stanford University, Stanford, California 94305, USA

[2]Geballe Laboratory for Advanced Materials, Stanford University, Stanford, California 94305, United States

[3]Stanford Institute for Materials and Energy Sciences, SLAC National Accelerator Laboratory, 2575 Sand Hill Road, Menlo Park, CA 94025, USA

[4]Laboratoire des Multimateriaux et Interfaces, Université de Lyon, 43 boulevard du 11 novembre 1918, 69622 Villeurbanne Cedex, France

[5]Institute of Organic Chemistry, Justus-Liebig University, Heinrich-Buff-Ring 17, 35392 Giessen, Germany

ABSTRACT

We demonstrate a new approach for engineering group IV semiconductor-based quantum photonic structures containing negatively charged silicon-vacancy (SiV⁻) color centers in diamond as quantum emitters. Hybrid SiC/diamond structures are realized by combining the growth of nano- and micro-diamonds on silicon carbide (3C or 4H polytype) substrates, with the subsequent use of these diamond crystals as a hard mask for pattern transfer. SiV⁻ color centers are incorporated in diamond during its synthesis from molecular diamond seeds (diamondoids), with no need for ion-implantation or annealing. We show that the same growth technique can be used to grow a diamond layer controllably doped with SiV⁻ on top of a high purity bulk diamond, in which we subsequently fabricate nanopillar arrays containing high quality SiV⁻ centers. Scanning confocal photoluminescence measurements reveal optically active SiV⁻ lines both at room temperature and low temperature (5 K) from all fabricated structures, and, in particular, very narrow linewidths and small inhomogeneous broadening of SiV⁻ lines from all-diamond nano-pillar arrays, which is a critical requirement for quantum computation. At low temperatures (5 K) we observe in these structures the signature typical of SiV⁻ centers in bulk diamond, consistent with a double lambda. These results indicate that high quality color centers can be incorporated into nanophotonic structures synthetically with properties equivalent to those in bulk diamond, thereby opening opportunities for applications in classical and quantum information processing.

TABLE OF CONTENT GRAPHIC

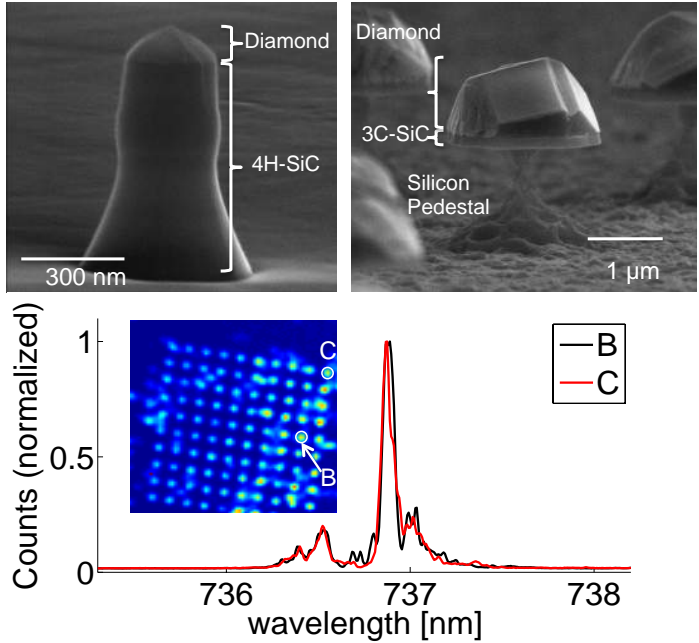

TEXT

Quantum emitters in solids are an important resource for diverse applications including quantum information processing (QIP) and sensing.[1-3] Some of the most common approaches include emitters such as InAs quantum dots and nitrogen-vacancy (NV$^-$) color centers that are incorporated into devices composed of a single material such as GaAs and diamond, respectively.[4,5] However, these approaches suffer from drawbacks including the limited coherence times of spins in InAs/GaAs quantum dots,[6] the necessity for operation at cryogenic temperatures,[4] incompatibility with Si CMOS platforms (as is needed for devices in optical interconnects), and large inhomogeneous broadening.[7] On the other hand, for NV$^-$ centers in diamond, drawbacks include a lack of a second order optical nonlinearity needed for efficient frequency conversion,[8] degradation of NV$^-$ center properties resulting from proximity of etched surfaces, and non-standard fabrication methods needed to carve optical structures in bulk diamond.[9,10] Hybrid solid-state schemes consisting of quantum emitters in individual nanocrystals coupled to semiconductor optical cavities in other materials have also been demonstrated;[11] however, they typically require AFM 'pick-and-place' assembly techniques[12] or complex film transfers.[13-16]

Silicon-vacancy (SiV$^-$) color centers in diamond have recently emerged as promising candidates for QIP applications.[17] Due to the inversion symmetry of the SiV$^-$ centers, they display superior spectral stability and narrow inhomogeneous broadening, with room temperature linewidths of a few nanometers and nearly transform-limited linewidths at low temperature.[17] Even more importantly, 70% of the photons are emitted into the strong zero-phonon line, compared to only 3% for NV$^-$ centers. Single photon generation from SiV$^-$ centers has by now been routinely demonstrated,[18,19] and their application in QIP, such as quantum key distribution (QKD), has been considered.[20] Moreover, this emitter offers a Λ-system energy level configuration, which is needed for all optical electron spin control, in weak or zero magnetic fields. This contrasts with the InAs/GaAs QDs that have to be charged with an extra electron or hole and placed in a strong magnetic field to realize a Λ-system.[21] Recently, optical initialization, readout, and coherent preparation of the electron spin of the SiV$^-$ center in bulk diamond was demonstrated.[22,23] However, as opposed to InAs/GaAs quantum dots, SiV$^-$s have yet to be incorporated into high quality nanophotonic structures without affecting their spectral signatures,[24] which will enable strong interaction and efficient interfacing with photons. The orbital relaxation limited spin coherence time of SiV$^-$ centers has been measured to be in the 10s of nanosecond range,[22,23] which is not yet competitive with the spin coherence time in charged InGaAs quantum dots.[25] However, there may be a possibility to improve the coherence by reducing the phonon density of states in the SiV$^-$ environment[26] via engineering nanostructures similar to those shown in this article.

Here we report a new approach for implementing nanophotonic structures with embedded diamond SiV⁻ centers of high quality, and apply it to two types of devices: diamond-SiC hybrid structures and all-diamond nanopillars arrays. Nanodiamonds and diamond films doped with SiV⁻ centers are grown from molecular diamond seeds (diamondoids) on SiC and high purity bulk diamond substrates, respectively. SiV⁻ color centers are incorporated in diamond during its synthesis, with no need for ion-implantation or annealing. Nanodiamonds grown from diamondoids in our approach are devoid of nitrogen vacancy centers, which could be present in the Ultra Dispersed Diamonds (UDD).[27] For hybrid diamond-SiC structures, functionalized nanodiamonds are used as a hard mask for pattern transfer into the 3C and 4H-SiC substrates, followed by subsequent removal of a buried Si sacrificial layer in the case of 3C-SiC. This simple technique results in high yield of devices containing SiV⁻ centers, avoiding the AFM 'pick-and-place' or film transfers. Moreover, combining two group IV semiconductors - diamond and silicon carbide - could potentially address challenges in each individual material, including difficulty in fabricating planar photonic structures in diamond, lack of doping for electrical devices, absence of second order optical nonlinearity in diamond,[28-31] and less developed quantum emitters in SiC.[32,33] In addition, diamond-SiC devices could benefit from the mature fabrication methods developed for SiC, which has produced high quality factor optical microcavities.[30,31] Finally, the method we describe in the article allows for positioning of diamond relative to SiC structures, as diamond itself is used as an etch mask. On the other hand, all-diamond monolithic nanopillar arrays are nanofabricated in SiV⁻ doped diamond films grown on bulk diamond substrates, using standard microfabrication methods (lithography and dry etching). The resulting nanopillars contain high quality SiV⁻ centers with very narrow linewidths and very low inhomogeneous broadening, thereby enabling implementation of quantum photonic devices containing identical quantum emitters, as needed for quantum simulations[34] and quantum networks.[35] As explained below, our growth methods allow us to control vertical positioning of SiV⁻ inside diamond films to ~10nm thickness at a chosen depth.

**Growth and fabrication of structures**

High quality nanodiamond crystals and diamond films are CVD grown on SiC and bulk diamond substrates, respectively, starting from diamondoid seeds covalently attached to a SiC or diamond surface (Fig. 1(a)). Diamondoids are a class of face-fused adamantane ($C_{10}H_{16}$) building blocks whose extension eventually leads to macroscopic diamond.[36-38] The lower diamondoids are adamantane ($C_{10}H_{16}$), diamantane ($C_{14}H_{20}$) and triamantane ($C_{18}H_{24}$), while the higher diamondoids begin with isomeric tetramantane ($C_{22}H_{28}$) and pentamantane ($C_{26}H_{32}$). While the lower diamondoids can readily be synthesized,[37,38] the higher diamondoids essentially are only accessible from petroleum sources via high pressure liquid chromatography (HPLC).[39] Unlike typical detonation diamond used to seed diamond CVD growth, diamondoids are free from nitrogen and graphitic impurities, with a precisely known molecular structure like most other small organic molecules. Diamondoids have been applied to diamond growth,[40-42] electron imaging[43] and electron emission devices.[44] Here, we covalently bond 7-dichlorophosphoryl[1(2,3)4]pentamantane[45] as a seed for high quality growth of fluorescent diamond nanoparticles and to form a bond between heteroepitaxial diamond layer and substrate.

The growth of the diamond is illustrated in Figure 1 (b). First, an oxide layer is generated on the substrate with exposure to oxygen plasma for 5 min at 400 mTorr pressure and 100 W power. For the devices presented in this paper, bulk 4H-SiC wafers (Cree) as well as heteroepitaxial 3C-SiC(100) thin films (~150 nm) grown on Si (100) via a standard two-step procedure were used.[46]

The sample is then soaked in toluene solution containing 1mM 7-dichlorophosphoryl[1(2,3)4]pentamantane. This process results in the generation of a covalently attached [1(2,3)4]pentamantane monolayer on the silicon carbide samples. From here, the sample is placed in a CVD reactor for the 'nucleation step' (gas mixture $H_2$: 5 sccm, $CH_4$: 10 sccm, Ar: 90 sccm, substrate temperature: 450 °C, microwave power: 300 W, pressure: 23 Torr) for ~20 min to enhance nucleation density observed from diamondoids. After nucleation, a 'growth step' (gas mixture $H_2$: 300 sccm $CH_4$: 3-7.5 sccm, substrate temperature: 830 °C, microwave power: 1300 W, pressure: 30 Torr, 1-2.5% $CH_4$ in $H_2$ carrier gas) is performed, with growth times calibrated to desired nanoparticle size. High quality diamond crystals with grain sizes ranging from 500 nm to 2 μm can be seen in the scanning electron microscope (SEM) images in Figures 1(c) and (d). As will be discussed below, in photoluminescence (PL) experiments we observe the presence of $SiV^-$ in the as-grown nanodiamonds. They were incorporated during the 'growth step' through diffusion of Si atoms from the plasma etching of SiC substrate, without need for subsequent annealing or ion implantation. Our growth approach does not degrade the quality of the transition lines of the $SiV^-$ center, and leads to properties comparable to $SiV^-$ in bulk diamond.[17] Although we don't have control over lateral positioning of $SiV^-$ centers in our growth approach, as opposed to ion implantation, we do have control over their vertical positioning, which enables us to grow $SiV^-$s only within ~10 nm thick diamond layer at a chosen depth.

The process flow for fabricating hybrid diamond–SiC devices is shown in Figure 2. First, high quality nano- and micro-diamond doped with $SiV^-$ was grown on 3C- and 4H-SiC as described above. The high quality of the seeding and growth process ensures optical quality of the fabricated structures, as will be demonstrated in the PL measurements below. Then, the diamond particle shape was transferred into the substrate through anisotropic etching of the substrate using the diamond crystals as the etch mask. Lastly, any additional processing (e.g., undercutting in case of 3C-SiC on Si substrate, or thinning in case of 4H-SiC) can be performed. As a proof of concept, we have applied this approach to generate two different types of hybrid diamond-SiC nanophotonic structures: (i) diamond - 4H-SiC nanowires (also applicable to 3C-SiC) and (ii) diamond-3C-SiC hemispherical microdome (akin to whispering gallery mode resonators). For the diamond - 4H-SiC nanowires, approximately 500 nm diameter diamond nanocrystals were first grown on 4H-SiC, and then used as hard mask to etch the underlying substrate using inductively coupled plasma (ICP) etching with $HBr/Cl_2$ chemistry, as illustrated in Figure 2(a). The SEM image of a diamond - 4H-SiC nanowires is shown in Figure 2(b). Alternatively, for fabricating the diamond-3C-SiC microdome structures, diamond microcrystals with diameters ~ 2 μm were first grown on 150 nm thick 3C-SiC epitaxial film on Si.[46] Then, the pattern of randomly distributed microdiamonds was transferred through 3C-SiC using the same etch recipe as above, followed by undercutting the sacrificial Si using the $XeF_2$ vapor phase silicon etcher, as illustrated in Figure 2(c). The SEM images of diamond - 3C-SiC microdomes are shown in Figure 2(d) and (e).

For monolithic nanopillar arrays, we grow a homoepitaxial layer of $SiV^-$ center doped thin film grown on diamond substrate using the same method as described earlier for the nanodiamonds. Starting with a bulk diamond substrate (Element Six, type Ib), a 70 nm diamond film containing $SiV^-$ was grown homoepitaxially via microwave plasma-assisted chemical vapor deposition (MPCVD). The $SiV^-$ centers are introduced through diffusion of Si atoms from plasma etching of the SiC placed near the diamond substrate during the growth. After the film growth, the nanopillars arrays are defined lithographically using evaporated gold as hard mask, as illustrated in Supplementary Figure 1(a). Figures 2(f) and (g) show SEM images of a typical fabricated array and an individual nanopillar.

**Optical characterization**

The presence of SiV⁻ centers in all described structures is confirmed by scanning confocal microscopy measurements. The custom made laser scanning confocal microscope consists of 532 nm continuous wave (CW) pump laser focused onto the sample through a high numerical aperture NA = 0.75 microscope objective, as shown in Figure 3(a). The photoluminescence (PL) is collected through the same objective and is sent into a single mode collection fiber with a dichroic mirror. A scanning galvanometer in the common path of the pump and collection scans the focal spot across the sample surface. This allows producing PL maps of the sample, as well as addressing individual devices using the scanning mirror. The collected emission is directed onto an avalanche photodiode (APD) for generating the PL map, or a high-resolution spectrometer for spectral characterization.

A typical laser scanning confocal microscope image for hybrid diamond-SiC nanowires is shown in Figure 3(b) (similar PL maps were obtained for nanodiamonds as well as SiC microdomes). The bright areas with high count rates indicate the presence of randomly distributed nanodiamonds containing SiV⁻ centers in described photonic structures. The room temperature PL spectrum of a typical diamond-SiC nanowire is presented in Figure 3(c) and exhibits a narrow emission peak at 738 nm, corresponding to the emission from SiV⁻ centers embedded in the diamond.[17] Similar spectra were observed prior to the fabrication of the photonic structures. Therefore, this confirms that the SiV⁻ centers in the nano- and micro-diamond crystals before the fabrication process are retained in the photonic structures post-fabrication. The low temperature PL spectra of SiV⁻s in nanodiamonds grown on SiC can be seen in Supplementary Figure 2. In this hybrid system at low temperature, SiV⁻s feature multiple emitter lines and strain induced spectral shifts in transitions, likely resulting from the lattice mismatch between the SiC substrate and diamond grown on top. Growth of such hybrid systems on substrates that have a better lattice match to diamond could be a route to addressing this issue, or simply a growth of all diamond homoepitaxial systems presented below.

In addition, a laser scanning confocal microscopy map of the diamond nanopillar array sample is shown in Figure 4(a), revealing strong PL at the locations of the nanopillars. High resolution spectra of the nanopillars (Figure 4(b)) confirm PL from SiV⁻ at room temperature. At low temperature (~5 K) the spectra (Figure 4(c) and (d)) show four distinct lines, consistent with the signature of the electronic energy levels of unstrained SiV⁻ centers in diamond – and so far observed in bulk diamond only.[17] Repeatable measurements of the spectrum of nanopillar A (as highlighted in Figure 4(a)), are presented in Figure 4(c) and clearly demonstrate the absence of bleaching over minutes timescale. Furthermore, different nanopillars in the same array (i.e., pillars B and C) show strong spectral overlap, as shown in Figure 4(d). Therefore, in addition to narrow linewidths, the demonstrated SiV⁻ centers incorporated in diamond nanopillar structures also feature small inhomogeneous broadening, potentially arising from the lower defect density and lower strain in the homoepitaxial diamond nanopillars (see Supporting Information). Although we do not observe inhomogeneous broadening of SiV⁻ lines at low temperature inside 250 nm all-diamond homoepitaxial nanopillars (within our spectrometer resolution limit of 4 GHz), we start to see some splitting in smaller, 130 nm diameter nanopillars (see Supplementary Figure 3). By counting the number of line quadruplets inside such structures we can estimate that on average we have ~5 SiV⁻s inside a 130 nm nanopillar, or doping density of $5\times10^{15}$ cm$^{-3}$. This agrees with our estimate of the density of SiV⁻s in hybrid diamond-SiC structures doped at similar level. Therefore, reduction in doping density is needed to isolate a single SiV⁻ inside such structures. As an additional confirmation of the fact that we have multiple SiV⁻s in our current structures, we have performed $g^{(2)}(t)$ measurements on the same 130 nm pillars shown in the Supplementary Figure 3. Antibunching is absent, as expected. The lifetime measurements on nanopillars reveals a mean lifetime of 1.18 ns (see Supplementary Figure 4), and both the PL linewidth and lifetime are comparable to that of single SiV$^-$ centers in bulk.[17]

The small inhomogeneous broadening in the larger pillars suggests significant potential for entangling different quantum emitters on a chip via photon interference, and for using such arrays to generate large entangled photon states efficiently - as needed for quantum networks and quantum simulations.[34,35] Therefore, our approach for engineering quantum photonic structures incorporating diamond SiV$^-$ not only enables the implementation of an efficient interface between a quantum emitter and a photon, but also does so without degradation of the properties of the quantum emitter itself. Moreover, the demonstrated nanopillar structures may also offer the opportunity to prolong the spin coherence time via reduced phonon density of states, thereby improving the potential of this system as a solid state quantum memory.

In conclusion, we have demonstrated a new approach for implementing nanophotonic structures with selectively incorporated high quality SiV$^-$ centers in nanodiamonds or diamond films during MPCVD growth (without need for ion implementation or annealing). We have successfully applied this approach to two types of structures: hybrid diamond-SiC structures (SiC microdomes and nanowires with diamond tips) and all-diamond nanopillar arrays with quantum emitters at tips of pillars. The hybrid diamond-SiC devices are fabricated by pattern transfer into the SiC substrate using diamond nanocrystal as hard mask, which simplifies fabrication, but also retains the SiV$^-$ centers post fabrication, as illustrated by the room temperature spectra. On the other hand, monolithic diamond nanopillar arrays were fabricated through a diamond thin film grown on high purity diamond substrate, followed by conventional microfabrication. In particular, we note the narrow linewidths, absence of bleaching, and small inhomogeneous broadening in the diamond nanopillars array at low temperature. This can open opportunities for entangling multiple quantum emitters on the chip. Future characterization will involve resonance fluorescence spectroscopy and Hong-Ou-Mandel interferometry. The developed structures could be employed as large arrays of individually addressable quantum emitters of indistinguishable photons for use in quantum networks. Future directions could also encompass utilizing the reduced phonon density of states to prolong the spin coherence time, in combination with the low strain environment, to improve the potential of this system as a solid state quantum memory.

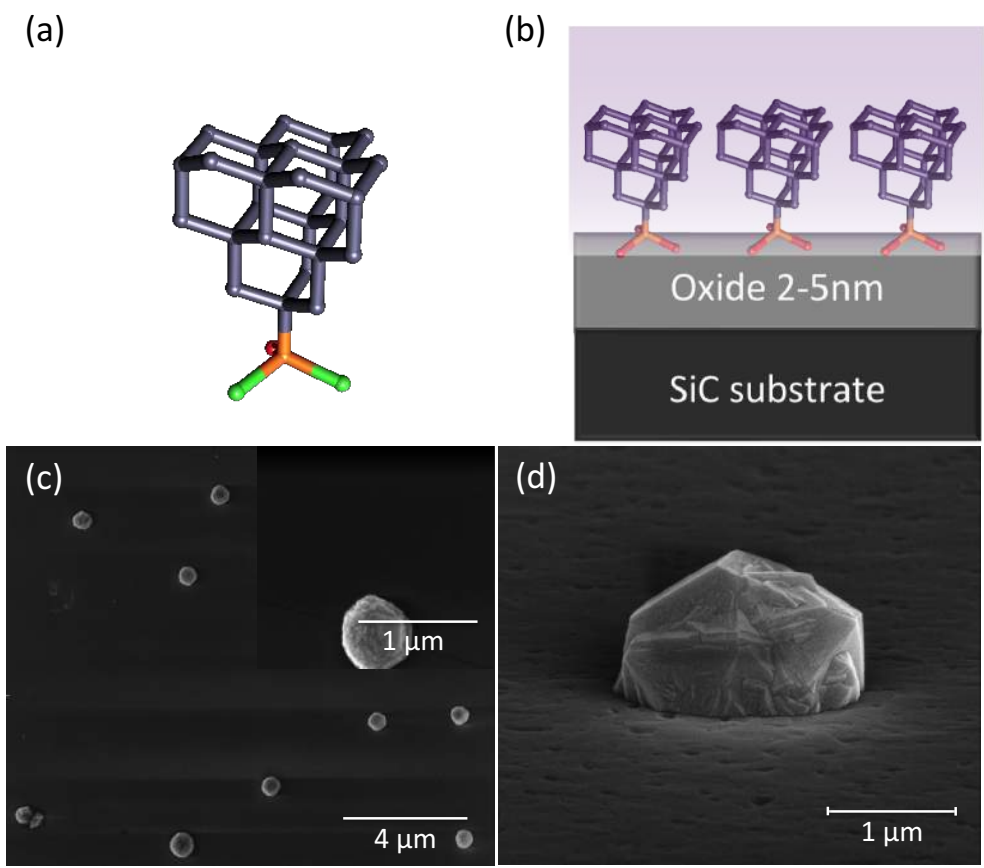

**Figure 1.** (a) Molecular structure of 7-dichlorophosphoryl[1(2,3)4]pentamantane. (b) Schematic of 7-dichlorophosphoryl[1(2,3)4]pentamantane on an oxide layer formed on top of SiC substrate. (c) Scanning electron micrograph (SEM) of 500 nm diameter nanodiamonds grown on 4H-SiC substrate. (d) SEM of a micrometer size diamond on 3C-SiC substrate.

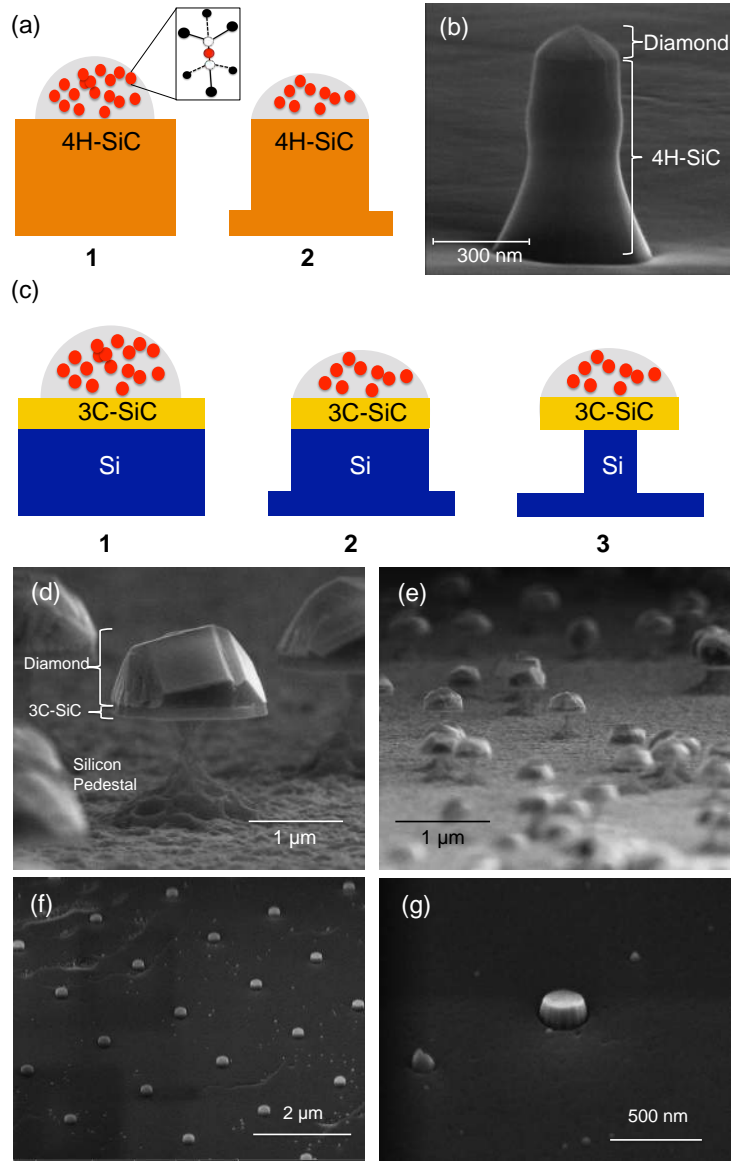

**Figure 2.** (a) Process flow for fabricating diamond – SiC nanowires through hard mask pattern transfer. The red dots represent the silicon vacancy centers in the diamond nanocrystals. (b) SEM image of a typical nanowire. (c) Process flow for hybrid diamond – SiC microdome structures fabricated through hard mask pattern transfer. (d) SEM image of a typical microdome structure. The high-quality diamond, as well as the heteroepitaxial interface are visible. (e) SEM image of an ensemble of microdome structures. (f) SEM image of a fabricated monolithic nanopillar array and (g) close-up SEM image of a monolithic nanopillar.

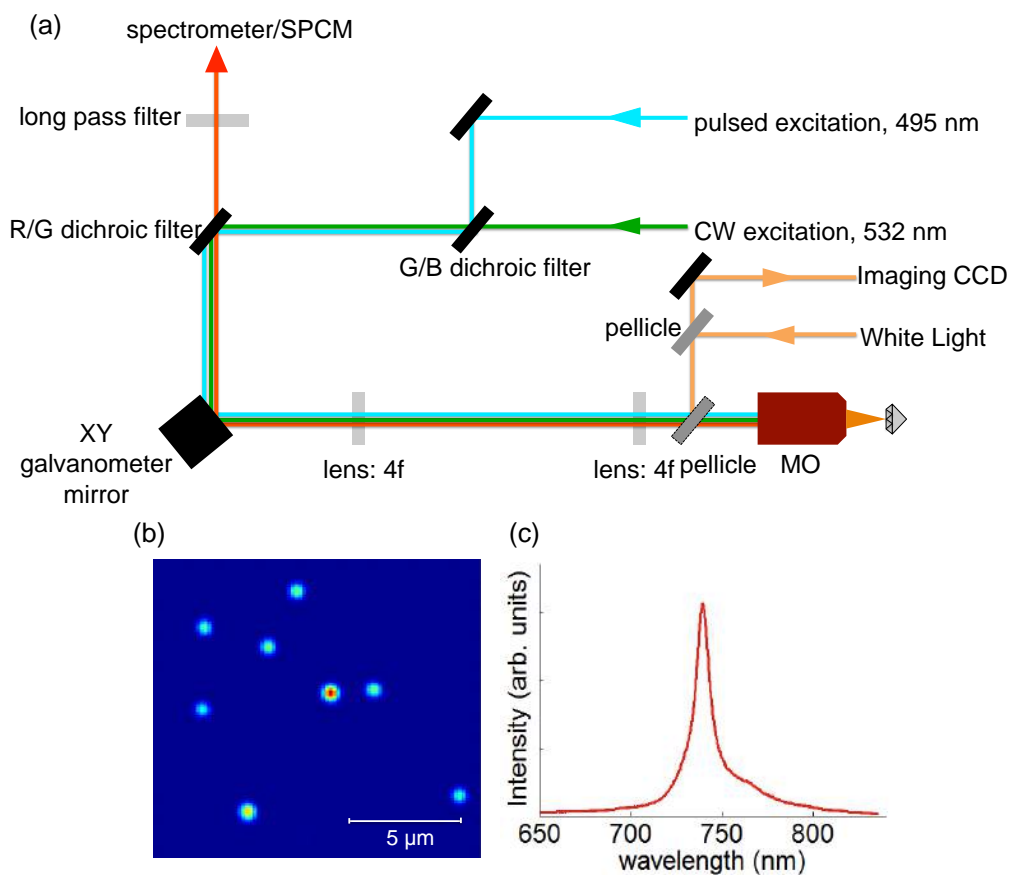

**Figure 3.** (a) Laser scanning confocal microscopy set-up for the photoluminescence measurement. (b) Laser scanning confocal microscope image of the diamond-SiC nanowires shown in Figure 2(b). This is representative of the scanning confocal microscope images for all the diamond nanocrystals and diamond-SiC (3C and 4H polytype) structures. (c) Typical PL spectrum of a diamond-SiC nanowire at room temperature, as observed on diamond nanocrystal or diamond-SiC hybrid structures. The PL signature at 738 nm is the emission signature of SiV⁻ centers.

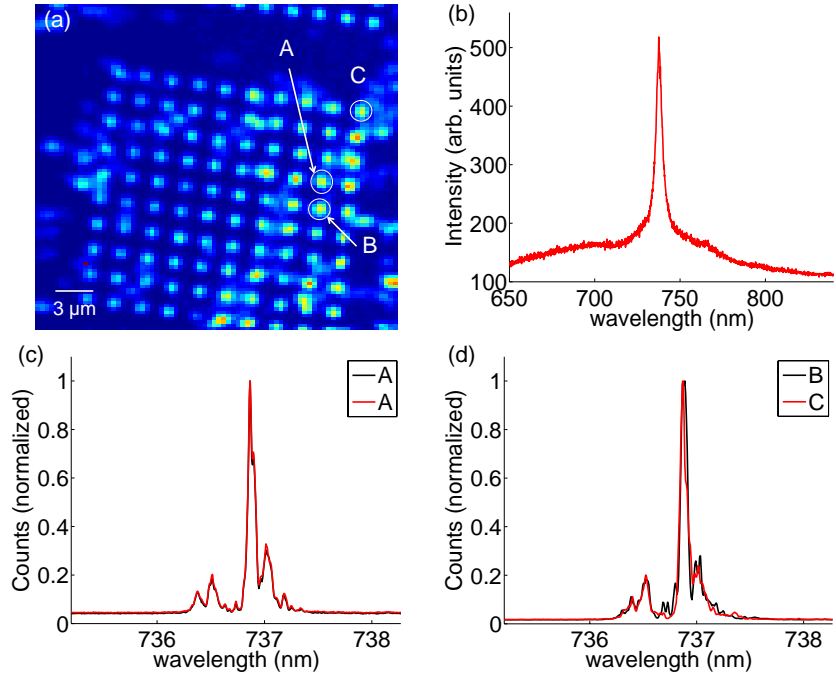

**Figure 4.** (a) Laser scanning confocal microscope image of a nanopillar array, with ~250 nm nanopillar diameter. (b) Room temperature PL spectrum of a typical nanopillar with ~250 nm diameter, with full-width-half-maximum (FWHM) linewidth of 7.4 nm. (c) Low temperature (~5 K) PL spectrum of the nanopillar labeled A in panel (a) containing $SiV^-$ center with 0.66 nm linewidth for the highest intensity transition. The black and red curves are repeated measurements of the spectra of nanopillar A, demonstrating the absence of collective emitter degradation and bleaching over timescale of minutes. (d) Low temperature PL spectra compared across nanopillars labeled B and C in panel (a). All spectra feature small inhomogeneous broadening and four transitions that are signatures of $SiV^-$.

## ASSOCIATED CONTENT

Supporting Information.

Process flow for fabricating diamond nanopillar arrays, low temperature PL of hybrid diamond-SiC systems, and spectral, lifetime, and $g^{(2)}(t)$ measurements on all-diamond nanopillars. This material is available free of charge via the Internet at http://pubs.acs.org.

## AUTHOR INFORMATION

### Corresponding Author

* Email: Jelena Vučković: jela@stanford.edu, Nicholas A. Melosh: nmelosh@stanford.edu, and Zhi-Xun Shen: zxshen@stanford.edu.

### Present Addresses

Thomas M. Babinec's present address: US Army Research Laboratory, 2800 Powder Mill Road, Adelphi MD 20783

### Author Contributions

The manuscript was written through contributions of all authors. All authors have given approval to the final version of the manuscript. ‡JLZ and HI contributed equally.

### Notes

The authors declare no competing financial interest.

## ACKNOWLEDGMENT

Financial support for materials synthesis is provided by the DOE Office of Basic Energy Sciences, Division of Materials Sciences through Stanford Institute for Materials and Energy Sciences (SIMES) under contract DE-AC02-76SF00515. Financial support for this work is also provided by Stanford Institute for Materials and Energy Sciences (SIMES), National Science Foundation grant DMR-1406028, AFOSR MURI on quantum metaphotonics and metamaterials, and the Deutsche Forschungsgemeinschaft (Schr 597/23-1). This work was performed in part at the Stanford Nanofabrication Facility of NNIN supported by the National Science Foundation under Grant No. ECS-9731293, and Stanford Nano Shared Facility. JLZ acknowledges support from the Stanford Graduate Fellowship. KM acknowledges support from the Alexander von Humboldt Foundation. LJZ thanks Yan-Kai Tzeng for the helpful discussions.

Supporting Information

**Fabrication of nanopillar arrays**

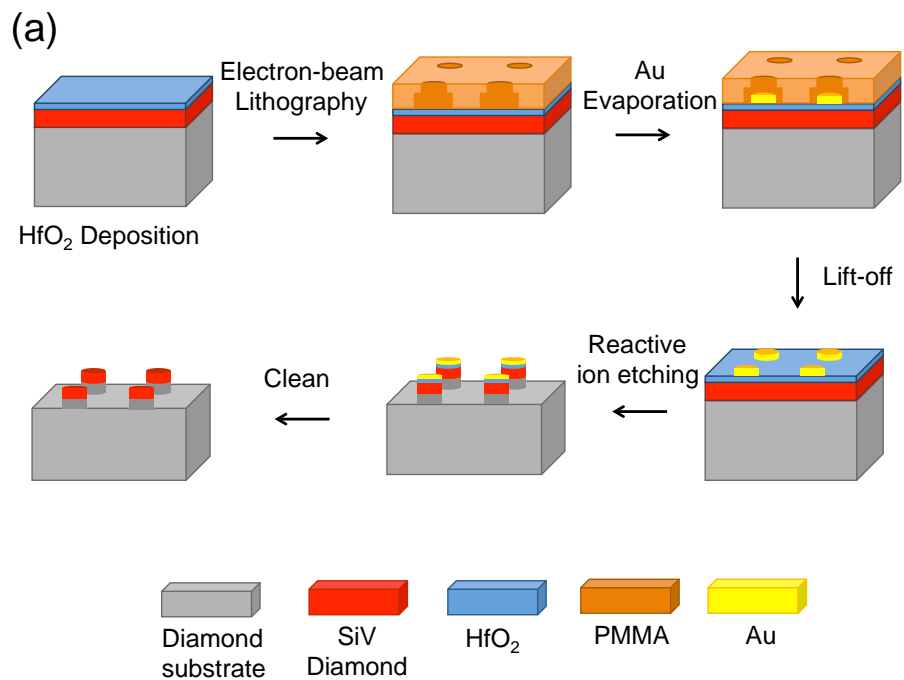

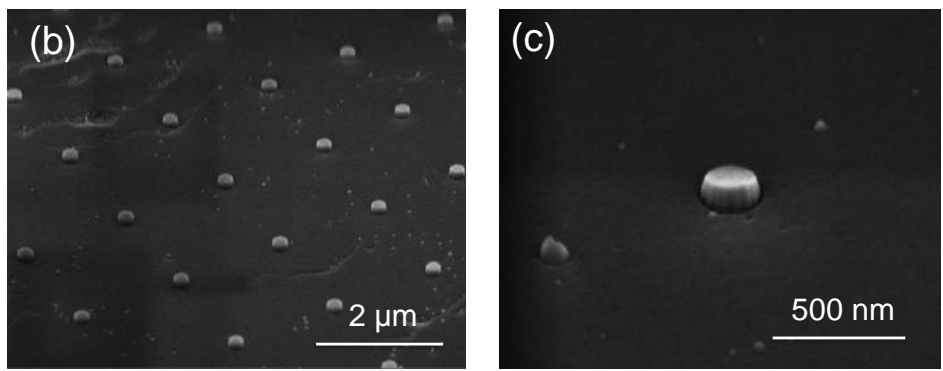

**Supplementary Figure 1.** (a) Process flow for fabricating diamond nanopillar arrays. Diamond film containing SiV⁻ is grown homoepitaxially on bulk diamond substrate (Element Six, type Ib). Subsequent ALD deposition of 5 nm $HfO_2$ facilitates metal adhesion. The nanopillar array is defined by electron-beam lithography in bilayer positive resist PMMA, followed by evaporation of gold as hard mask and the lift-off of PMMA. Lastly, the nanopillar array pattern is transferred into the diamond substrate past the SiV⁻ layer so that the background is SiV⁻ free, and the gold and $HfO_2$ are cleaned off. (b) SEM image of a fabricated nanopillar array and (c) close-up SEM image of a nanopillar.

To fabricate the nanopillars arrays, 5 nm HfO$_2$ was deposited as an adhesion layer via Atomic Layer Deposition (ALD) following the diamond film growth. Arrays of nanopillars with diameter 115–250 nm diameter and 150 nm height were defined using electron-beam lithography followed by ICP RIE, with electron-beam evaporated gold as hard mask. Lastly, the gold hard mask and adhesion layer were removed using gold etch and piranha clean recipes, leaving nanopillars with silicon vacancy centers near the top of the nanopillars.

**Low temperature PL of hybrid diamond-SiC systems**

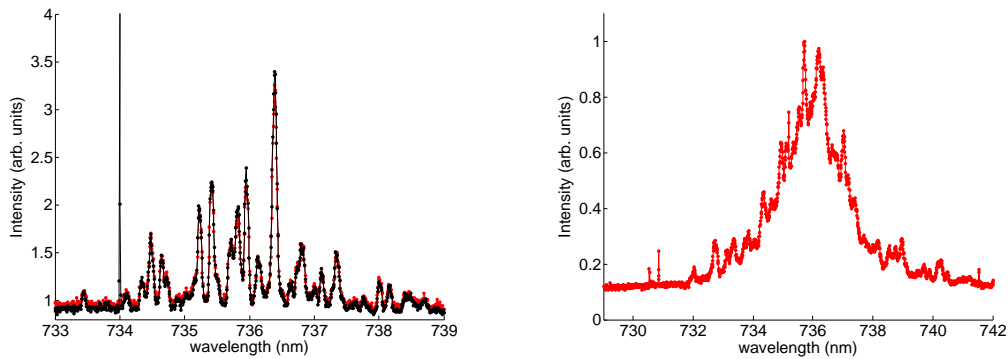

**Supplementary Figure 2**: (Left) Low temperature PL emission spectrum of a 200 nm nanodiamond on 4H-SiC. (Right) Low temperature PL emission spectrum of a diamond-SiC nanowire as described in the manuscript. In this hybrid system at low temperature, SiV⁻s feature multiple emitter lines and strain induced spectral shifts in transitions, likely resulting from the lattice mismatch between the SiC substrate and diamond grown on top.

As can be seen from the low temperature spectra of nanodiamonds on SiC (Figure S2), SiV⁻s in these hybrid systems feature multiple emitter lines and strain induced spectral shifting at low temperature. The left figure shows the low temperature PL spectrum of a 200 nm nanodiamond on 4H-SiC, while the right figure shows the low temperature PL spectrum of the diamond-SiC nanowires in the manuscript. Therefore, homoepitaxial system results in much smaller strain and smaller inhomogenous broadening.

**Doping density**

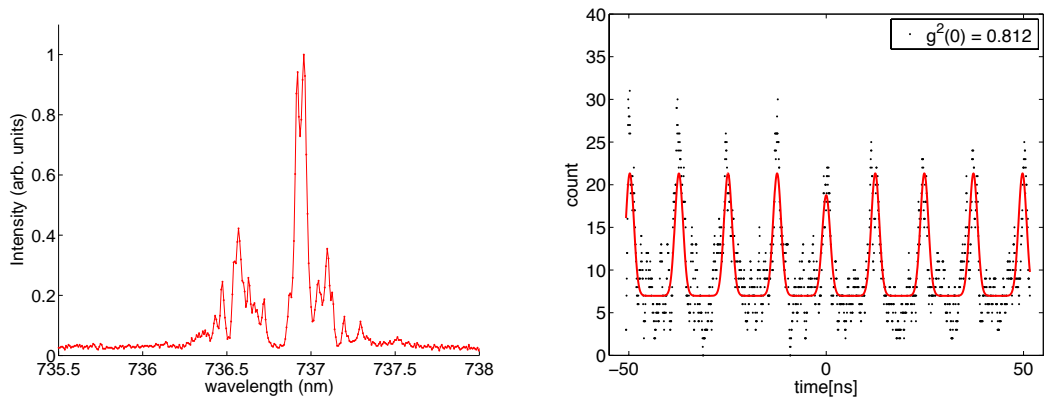

**Supplementary Figure 3**: Figure R2: (Left) Low temperature PL emission spectrum of a nanopillar 130 nm in diameter. (Right) Representative $g^{(2)}(t)$ measurements on the smallest nanopillars produced with ~130 nm diameter. $g^{(2)}(0)$~0.81 is estimated from the fit, indicating multiple emitters inside the pillar, as expected from the number of line quadruplets shown in the left figure.

Since we used the same Si doping densities both for the heteroepitaxial and homoepitaxial growth process, we can use strain split spectra of heteroepitaxial structures to estimate the number of SiV⁻s per pillar. The shown spectra of diamond grown on SiC suggest multiple (2+) emitters per nanodiamond of diameter 200 nm. In the nanopillar structures, we see no strain splitting in 250 nm diameter pillars, as shown in the main text, but strain splitting starts to manifest in 130 nm pillars, as shown in Supplementary Figure 3(left). One emitter per pillar (~ 130 nm diameter) corresponds to a SiV⁻ density of $1.0 \times 10^{15}$ cm$^{-3}$, so we expect an emitter density of several $10^{15}$ cm$^{-3}$. This is consistent with the density estimated from the ~200 nm diameter nanodiamond.

We have then performed $g^{(2)}(t)$ measurements on the all diamond, homoepitaxial nanopillars with the smallest diameters (~ 130nm diameter), as shown in Supplementary Figure 3(right), but didn't observe anti-bunching. This is in agreement with estimated SiV⁻ density above.

However, multiple emitters with small inhomogeneous broadening embedded in photonic structures are also very interesting for multi-emitter CQED, which is not possible with quantum dots (as a result of large inhomogeneous broadening). This is in fact another topic of our research, as in such multi-emitter system one can increase Rabi splitting in proportion to $\sqrt{N}$, where N is the number of emitters. This increases the operating speed of the devices based on the strong coupling CQED effects, and makes reaching the strong coupling regime easier.

**Inhomogeneous broadening of multiple SiV⁻s**

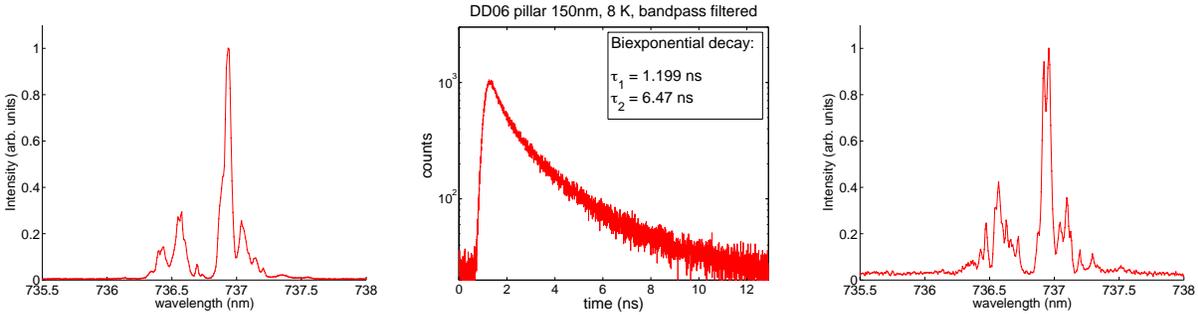

**Supplementary Figure 4:** (Left) PL spectrum of a nanopillar 150 nm in diameter. The strongest transition has a FWHM of 28.3 GHz (or 0.0513 nm, 117 μeV). (Center) At the same spot, the lifetime was measured to be 1.20 ns, where $\tau_1$ corresponds to the lifetime of SiV⁻, and $\tau_2$ corrresponds to the lifetime of the background signal. (Right) PL spectrum of a nanopillar 130 nm in diameter, showing strain induced spectral shifting.

The linewidth of all presented PL measurements was limited by the spectrometer resolution of 4.04 GHz, and is therefore greater than or equal to the true inhomogeneously broadened linewidth measured with other methods such as photoluminescence excitation.

We measured the spectra and lifetimes at the same spots in nanopillars (diameter ~ 130 - 150 nm), an example of which is shown in Supplementary Figure 4 (left, center). In the left figure, the spectrum is taken on a nanopillar with 150 nm diameter, while the strongest transition has a FWHM of 28.3 GHz (or 0.0513 nm, 117 μeV). At the same spot, the lifetime was measured to be 1.20 ns as shown in the center figure, comparable to that of typical SiV⁻ centers in bulk. Measurement of the lifetime in eight nanopillars reveals a mean lifetime of 1.177 ns with a standard deviation of 0.064 ns. Both the PL linewidth and lifetime are comparable to that of single SiV⁻ centers in bulk [*Nat. Commun.* 5:4739 (2014)], and the spectral shifting between pillars is within the linewidth of individual transitions, as shown in Figure 4(d) in the manuscript. It should be noted that we see no strain induced spectral shifting in 250 nm pillars, but we did start to see some spectral shifting in 130 nm pillars, as shown in Supplementary Figure 4(right).